\documentstyle[12pt,aaspp4,flushrt]{article}

\begin{document}
\title{Tracing the Universe with Clusters of Galaxies}
\author{Neta A. Bahcall \\
Princeton University Observatory}

\section*{Abstract}
  
Clusters of galaxies, the most massive virialized systems known, provide a  
powerful tool for studying the structure, the mass density, and the  
cosmology of our universe. Clusters furnish one of the best estimates of the  
dynamical mass density parameter on 1 Mpc scale, $\Omega _{dyn}$; the best  
measure of the baryon density fraction in the universe; an excellent tracer  
of the large-scale structure of the universe; and, most recently, a powerful  
tracer of the evolution of structure in the universe and its unique  
cosmological implications. I review the above measures and show that they  
portray a consistent picture of the universe and place stringent constraints  
on cosmology. Each of these independent measures suggests a low-density  
universe, $\Omega _{m}$ $\simeq $ 0.3 $\pm $ 0.1, with mass approximately  
tracing light on large scales.  
  
\section{Cluster Masses and $\Omega _{m}$}  
  
Masses of clusters of galaxies can be determined (within a given radius)  
from three independent methods: 1) in the optical, from velocity dispersion  
of galaxies in the clusters, assuming hydrostatic equilibrium (e.g.  
, Zwicky, 1957, Bahcall, 1977, Peebles 1980, Carlberg, et al.,  
1996); 2) in X-rays, from the temperature of the hot intracluster gas (  
e.g., Jones and Forman, 1984, Sarazin, 1986, Hughes, 1989, Evrard,  
et al., 1996); and 3) from lensing, using weak gravitational lens  
distortions of background galaxies caused by the intervening cluster mass  
(e.g., Tyson, et al., 1990, Kaiser and Squire, 1993, Smail,   
et al., 1995, Kneib, this volume). All three independent methods yield  
consistent results for the mass of rich clusters (typically measured within $%
R$ $\simeq $ $0.5-1.5h^{-1}$ $Mpc$); (e.g., Lubin and Bahcall 1993,  
Bahcall 1995, Fischer and Tyson 1997, Hjorth, et al., this volume,  
and references therein). The rms scatter in the mass determination among the  
different methods is typically $\sim \pm \;30\%$, with no significant bias.  
The observed consistency among the different methods ensures that we can  
reliably determine cluster masses, within the observed scatter. The masses  
of rich clusters range from $\sim 10^{14}$ to $\sim $ $10^{15}h^{-1}M_{\odot  
}$ within $1.5h^{-1}$ $Mpc$ radius of the cluster center. When normalized by  
the cluster luminosity, a median value of $M/L_{B}$ $\simeq $ $300$ $\pm $ $%
100h$ is observed for rich clusters, independent of the cluster luminosity,  
velocity dispersion, or other parameters. (Here $L_{B}$ is the total blue  
luminosity of the cluster, corrected for internal and Galactic absorption).  
(See also Carlberg, et al., this volume, for detailed results of  
the CNOC cluster survey.) This mass-to-light ratio, when integrated over the  
luminosity density of the universe, yields a dynamical mass density of $%
\Omega _{dyn}$ $\simeq $ $0.2$ on $\sim $ $1.5h^{-1}$ $Mpc$ scale. This  
density assumes that all galaxies (and other large structures) exhibit the  
same high $M/L_{B}$ ratio. If the mass distribution is not biased, 
i.e., if mass follows light on large scales, then a global $\Omega _{m}$ $%
\simeq $ $0.2$ is implied for the cosmological density parameter. If, on the
other hand, as desired by theoretical considerations, the universe has a
critical density ($\Omega _{m}$ $=1$), than most of the mass of the universe
has to be unassociated with galaxies (i.e., with light, even to
these large scales of $\sim $ $1.5h^{-1}Mpc$), and reside, instead, mostly
in the ``voids'' where there is no light. An $\Omega _{m}$ $=1$ universe
thus requires a large bias in the distribution of mass versus light, with
mass distributed considerably more diffusely than light.

Is there a strong bias in the universe, with most of the dark matter
residing on large scales, well beyond galaxies and clusters? A recent
analysis of the mass-to-light ratio of galaxies, groups and clusters
(Bahcall, Lubin and Dorman, 1995) suggests a negative reply. The study shows
that while the M/L ratio of galaxies increases with scale up to radii of $R$ 
$\sim 0.1-0.2h^{-1}Mpc$, due to large dark halos around galaxies, this ratio
appears to flatten and remain approximately constant for groups and rich
clusters, to scales of $\sim $ $1.5Mpc$, and possibly even to the larger
scales of superclusters (Fig. 1). The flattening occurs at $M/L_{B}\simeq
200-300h$, corresponding to $\Omega _{m}$ $\simeq $ $0.2$. This observation
is contrary to the classical picture where the relative amount of dark
matter is believed to increase with scale (possibly reaching $\Omega _{m}$ $%
=1$ on large scales). The present data suggest that most of the dark matter
may be associated with the dark halos of galaxies and that clusters do not
contain a substantial amount of additional dark matter, other than that
associated with (or torn-off from) the galaxy halos, and the hot
intracluster gas. This flattening of M/L with scale, if confirmed by further
larger-scale observations, suggests that the relative amount of dark matter
does not increase significantly with scale (above $\sim 0.2h^{-1}Mpc$); in
that case, the mass density of the universe is low, $\Omega _{m}$ $\simeq $ $%
0.2-0.3$, with no significant bias.

\section{Baryons in Clusters}

Clusters of galaxies contain many baryons, observed as gas and stars. Within 
$1.5h^{-1}Mpc$ of a rich cluster, the X-ray emitting gas contributes $\sim $
$3-10h^{-1.5}\%$ of the cluster virial mass (or $\sim $ $10-30\%$ for $h=%
\frac{1}{2}$ (Briel, et al., 1992, White and Fabian, 1995). Visible
stars contribute another $\sim $ $5\%$. Standard Big-Bang nucleosynthesis
limits the baryon density of the universe to $\Omega _{b}$ $\simeq
0.015h^{-2}$ (Walker, et al., 1991). This suggests that the baryon
fraction in rich clusters exceeds that of an $\Omega _{m}=1$ universe by a
factor of $\sim 3$ (White, et al., 1993, Lubin, et al.,
1996). Detailed hydrodynamic simulations (above references) show that
baryons are not preferentially segregated into rich clusters. This implies
that either the mean density of the universe is considerably smaller, by a
factor of $\sim 3$, than the critical density, or that the baryon density of
the universe is much larger than predicted by nucleosynthesis. The observed
baryonic mass fraction in rich clusters, combined with the nucleosynthesis
limit (Walker, et al., 1991, Tytler, this volume), suggest $\Omega
_{m}$ $\simeq $ $0.2-0.3$; this estimate is consistent with $\Omega _{dyn}$ $%
\simeq $ $0.2$ determined from cluster dynamics.

\section{The Cluster Mass Function}

The observed mass function (MF) of clusters of galaxies, $n(>M)$, describes
the number density of clusters above a threshold mass $M$; it serves as a
critical test of theories of structure formation in the universe. The
richest, most massive clusters are thought to form from rare high peaks in
the initial mass-density fluctuations; poorer clusters and groups form from
smaller, more common fluctuations. Bahcall and Cen (1993) determined the MF
of clusters of galaxies using both optical and X-ray observations of
clusters. They compared the observed mass function of galaxy clusters with
predictions of N-body cosmological simulations (Bahcall and Cen 1992) for
standard ($\Omega _{m}=1$) and low-density ($\Omega _{m}<$ $1$; flat or
open) Cold Dark Matter (CDM) models. They find that standard $\Omega _{m}$ $%
=1$ CDM models, with any normalization, can not reproduce well the observed
cluster mass function. An $\Omega _{m}$ $=1$ model requires a low
normalization, $\sigma _{8}$ $\simeq $ $0.5$ (where $\sigma _{8}$ is the rms
mass fluctuation amplitude on $8h^{-1}Mpc$ scale), in order to fit the
observed abundance of richness class $\gtrsim 1$ clusters. (A $\sigma _{8}$ $%
\simeq 1\;\Omega _{m}\simeq 1$ model overproduces massive clusters by an
order of magnitude.) This low $\sigma _{8}$ (high bias) model, however, is
too steep at the most massive tail of the cluster MF. A low density ($\Omega
_{m}$ $\simeq $ $0.2-0.3$) CDM model, flat or open, with little or no bias,
provides a good fit to the data.

The present-day cluster abundance (from the mass or temperature functions of
clusters) places one of the strongest constraints on cosmology (see above;
also White, et al., 1993, Eke, et al., 1996, Viana and
Liddle, 1996, Pen, 1997); $\sigma _{8}$ $\Omega _{8}^{0.5}$ $\simeq $ $%
0.5\pm 0.05$ (with a small dependence on $\Lambda $). This constraint, while
powerful, is degenerate in $\Omega _{m}$ and $\sigma _{8}$; it requires that
models with $\Omega _{m}$ $=1$ have low normalization ($\sigma _{8}$ $\sim $ 
$0.5$), i.e., strongly biased models, with mass distributed
considerably more diffusely than light, since $\sigma _{8}(gal)$ $\simeq 1$
is observed for optical galaxies); models with $\Omega _{m}$ $\simeq 0.25$
need to have a high normalization ($\sigma _{8}$ $\simeq $ $1$, comparable
to optical galaxies, i.e., an unbiased universe with mass tracing
light on large scales). (The cluster mass function of Bahcall and Cen, 1993
(see above) best fits the $\Omega _{m}$ $\simeq $ $0.25$ unbiased models,
especially at the high-mass end.) Standard $\Omega _{m}$ $=1$ CDM models
with COBE normalization of $\sigma _{8}$ $\simeq $ $1$ are strongly ruled
out by the observed present-day cluster abundance; they produce too many
massive clusters.

\section{Evolution of the Cluster Abundance}

The observed present-day abundance of rich clusters places one of the
strongest constraints on cosmology (\S 4): $\sigma _{8}$ $\Omega _{m}^{0.5}$ 
$\simeq $ $0.5\pm 0.05$. This constraint is degenerate in $\Omega _{m}$and $%
\sigma _{8}$: models with $\Omega _{m}=1$ and $\sigma _{8}$ $\simeq $ $0.5$
are indistinguishable from models with $\Omega _{m}\simeq 0.25$ and $\sigma
_{8}$ $\simeq $ $1$ (Eke, et al. 1996; Pen 1997; Kitayama and Suto
1997). In a recent paper (Bahcall, Fan, and Cen 1997) we show that the 
{\em evolution} of the cluster abundance as a function of redshift breaks
the degeneracy between $\Omega _{m}$and $\sigma _{8}$ and determines each of
the parameters independently (see also Carlberg, et al., this
volume). The growth of high-mass clusters depends strongly on the cosmology $%
-$ mainly $\sigma _{8}$ and $\Omega _{m}$ (e.g., Press and
Schechter, 1974, Peebles, 1993, Cen and Ostriker, 1994, Jing and Fang, 1996,
Eke, et al., 1996, Viana and Liddle, 1996, Oukbir and Blanchard,
1997). In low-density (high-$\sigma _{8}$) models, density fluctuations
evolve and freeze out at early times, thus producing only little evolution
at recent times ($z<1$). In an $\Omega _{m}$ $=1$ universe, the fluctuations
start growing more recently thereby producing strong evolution in recent
times: a large increase in the abundance of massive clusters is expected
from $z\simeq $ $1$ to $z\simeq 0$. The evolution is so strong that finding
even a few Coma-like clusters at $z>0.5$ over $\sim $ $10^{3}deg^{2}$ of sky
would contradict an $\Omega _{m}=1$,$\;\sigma _{8}$ $\simeq $ $0.5$ model,
where only $\sim $ $10^{-2}$ such clusters would be expected.

The evolution of the abundance of massive (Coma-like) clusters with mass $M$ 
$(\leq $ $R_{phy}=1.5$ $h^{-1}Mpc)\geq $ $6.3\times 10^{14}h^{-1}M_{0}$ is
presented in Fig. 2 (from Bahcall, et al., 1997). The expected
evolution is obtained from large-scale cosmological N-body simulations ($%
400h^{-1}Mpc$ box size) for different cosmologies: Standard Cold Dark Matter
(SCDM, $\Omega _{m}$ $=1$, COBE normalized: $\sigma _{8}$ $=$ $1.05$), SCDM
with $\sigma _{8}=$ $0.53$ (normalized to the present-day cluster abundance;
(\S 4), Mixed Dark Matter (MDM: hot + cold), Open CDM (OCDM, $\Omega
_{m}=0.35$, $\sigma _{8}=0.8$), and Lambda CDM (LCDM, $\Omega _{m}$ $
=0.4,\Lambda =0.6,\sigma _{8}=0.8$) (see Bahcall, et al., 1997 for
details). Several effects are clearly seen in Fig. 2:

\begin{enumerate}
\item  The evolution of the abundance of high-mass clusters breaks the
degeneracy between $\Omega _{m}$ and $\sigma _{8}$ that exists at $z$ $%
\simeq $ $0$;

\item  Low - $\sigma _{8}$ models (high $\Omega _{m}$) evolve much faster
than high - $\sigma _{8}$ models (low $\Omega _{m}$). The abundance of
clusters with this mass decreases by a factor of $\sim 10^{2}$ from $z=0$ to 
$z\simeq 0.5$ for biased ($\sigma _{8}\simeq $ $0.5-0.6$) SCDM models, while
the decrease is much shallower, only a factor of $\sim 5$, for the $\Omega
_{m}$ $\simeq $ $0.3(\sigma _{8}$ $\simeq 0.8\ )\;$models.

\item  Even at reasonably nearby redshifts of $z\sim 0.5$, the difference in
cluster abundance between high and low $\Omega _{m}$ models (low and high $%
\sigma _{8}$) is very high (factor of $\sim $ $10^{2}$) for these high mass
clusters.

\item  The available data, using the CNOC cluster survey (Carlberg, 
et al., 1997) and the distant EMSS survey (Luppino and Gioia, 1994), are
consistent with the low-density models (OCDM, LCDM), and inconsistent with
the biased $\Omega _{m}$ $=1$ CDM models (see Fig. 2, and Bahcall,
et al., 1997). Too many high mass clusters are observed at $z\simeq 0.5-0.8$
to be consistent with $\Omega _{m}$ $=1$, $\sigma _{8}$ $\simeq $ $0.5$
models; these biased models predict a factor of $\sim 10^{2}$ less clusters
than observed.
\end{enumerate}

The mild observed evolution of cluster abundance places a powerful
constraint on cosmology by breaking the degeneracy between $\Omega _{m}$ and 
$\sigma _{8}$; we find $\sigma _{8}$ $=0.83\pm 0.15$ and $\Omega _{m}$ $%
\simeq $ $0.3\pm 0.1$. (See also Carlberg, et al., this volume, for
similar results). We present these constraints in Fig. 3.

Fan, Bahcall and Cen (1997) investigate the reason for the strong
(exponential) dependence of the evolution rate on $\sigma _{8}$; they show
that $\Omega _{m}$ is determined mostly from the {\em normalization} of
the cluster abundance, while $\sigma _{8}$ is determined mostly from the 
{\em rate} of evolution. The exponential dependence of the evolution rate
on $\sigma _{8}$ arises because clusters of a given mass represent rarer
density peaks in low $\sigma _{8}$ models compared with high $\sigma _{8}$
models, thus evolving considerably faster in low $\sigma _{8}$ Gaussian
models. The high $\sigma _{8}$ value required by the mild observed cluster
evolution rate implies a bias parameter of $b$ $\simeq $ $\sigma
_{8}^{-1}\simeq $ $1.2$ $\pm 0.2$, i.e., a nearly unbiased universe with
mass approximately following light on large scale. The $\Omega _{m}\simeq
0.3 $ $\pm $ $0.1$ constraint, obtained from the cluster abundance
normalization for this $\sigma _{8}$ value, agrees well with the 
{\em independent} constraints placed by cluster dynamics and by the high baryon
fraction observed in clusters (\S 2-3).

\section{Clusters and Large-Scale Structure}

Clusters of galaxies serve as excellent tracers of the large-scale structure
of the universe. The strong cluster correlation function (Bahcall and
Soneira 1983, Klypin and Kopylov 1983) was the first to reveal the common
existence in the universe of large-scale structures to $\sim 50h^{-1}Mpc$ or
more. The richness-dependence of the cluster correlation function (Bahcall
and Soneira 1983, Bahcall and West 1992), is now observed for various
samples of clusters (including the APM survey (Croft, et al.,
1997), with consistent results and a weakening of the richness dependence at
the richest cluster tail). The cluster correlation function has been used
successfully in constraining cosmological models. The strong amplitude and
large-scale extent of the cluster correlations are inconsistent with $\Omega
_{m}$ $=1$ CDM models, for any normalization; the correlations are in good
agreement with low-density, $\Omega _{m}$ $\sim $ $0.2-0.3\;$CDM models
(Bahcall and Cen 1992, Croft, et al., 1997). Other large-scale
structure observations such as the power-spectrum (of galaxies and
clusters), the galaxy correlation function, and the cluster peculiar
motions, all suggest a low-density universe, with $\Omega _{m}h\sim 0.2$
(e.g., Maddox, et al., 1990, Peacock and Dodds 1992, Bahcall and Oh
1996, Croft, et al.,1997, Tadros, et al., 1997, Einasto, 
et al., 1997 and this volume, and Retzlaff , et al., 1997
and this volume). New large-scale surveys of galaxies and clusters,
currently underway, will further improve the constraint placed on this
parameter

\section{Is $\Omega _{m}<1$ ?}

Much of the observational evidence from clusters and from large-scale
structure suggests that the mass density of the universe is sub-critical: $%
\Omega _{m}\simeq 0.2-0.3$. I summarize these results below:

\begin{enumerate}
\item  The masses and the M/L(R) relation of galaxies, groups, and clusters
suggest $\Omega _{m}\simeq 0.2-0.3$(\S 2).

\item  The high baryon fraction in clusters of galaxies suggests $\Omega
_{m} $ $\simeq $ $0.2-0.3$ (\S 3).

\item  Various observations of large-scale structure (the cluster mass
function, the correlation function, the power spectrum, and the peculiar
velocities on Mpc scale) all suggest $\Omega _{m}h$ $\simeq 0.2$ (for a
CDM-type spectrum (\S\ 4,6).

\item  The evolution of the cluster abundance to $z\simeq 0.5-1$ yields $%
\Omega _{m}$ $\simeq 0.$3$\pm 0.1$ (\S 5).

\item  All the above independent observations suggest a low-density
universe, $\Omega _{m}$ $\simeq $ $0.3$ $\pm $ $0.1$.

\item  If $H_{0}\sim $ $70kms^{-1}Mpc^{-1}$, as indicated by a number of
recent observations, then the observed age of the oldest stars requires $%
\Omega _{m}$ $\ll $ $1$.

\item  Peculiar motions on large scales are too uncertain at the present
time (suggesting density values that range from $\Omega _{m}$ $\sim $ $0.2$
to $\sim $ $1$) to reliably constrain $\Omega _{m}$. Future results, based
on larger and more accurate surveys, will help constrain this parameter
(see, e.g., Strauss and Willick, 1996; Dekel; Gramann; this volume).
\end{enumerate}

I thank Volker Mueller and the organizing committee of the Potsdam Cosmology
Workshop for their hospitality and for an excellent, productive, and fun
meeting. This work is supported by NSF grant AST93-15368.

\newpage
\begin{figure}

\epsfysize=600pt \epsfbox{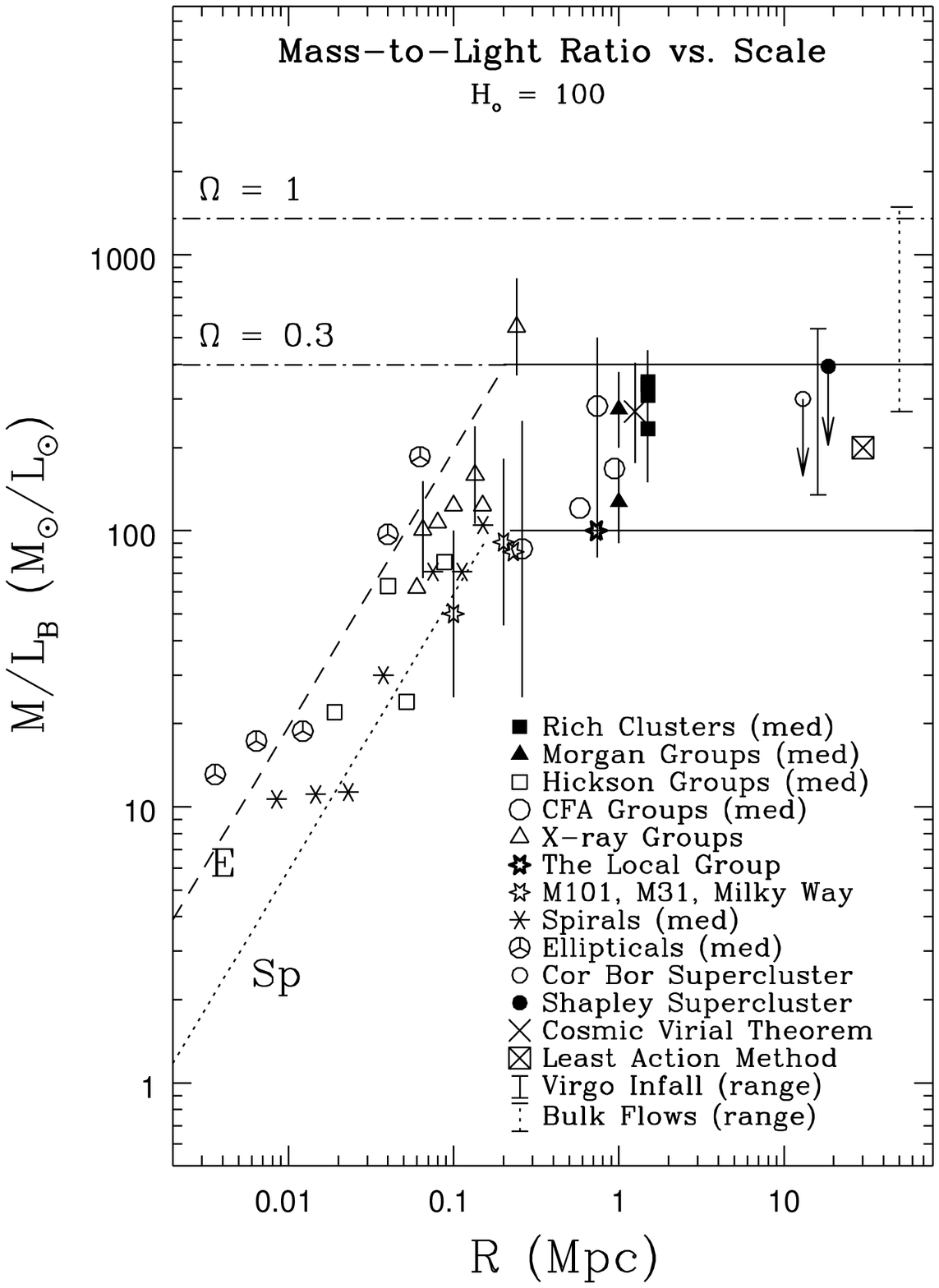}

\vspace{1cm}
Fig. 1 The dependence of $M/L$ on scale, $R$, for galaxies, groups,
clusters, and larger-scale structures. From Bahcall, Lubin and Dorman (1995).
\end{figure}
\begin{figure}
\vspace{-4cm}

\epsfysize=600pt \epsfbox{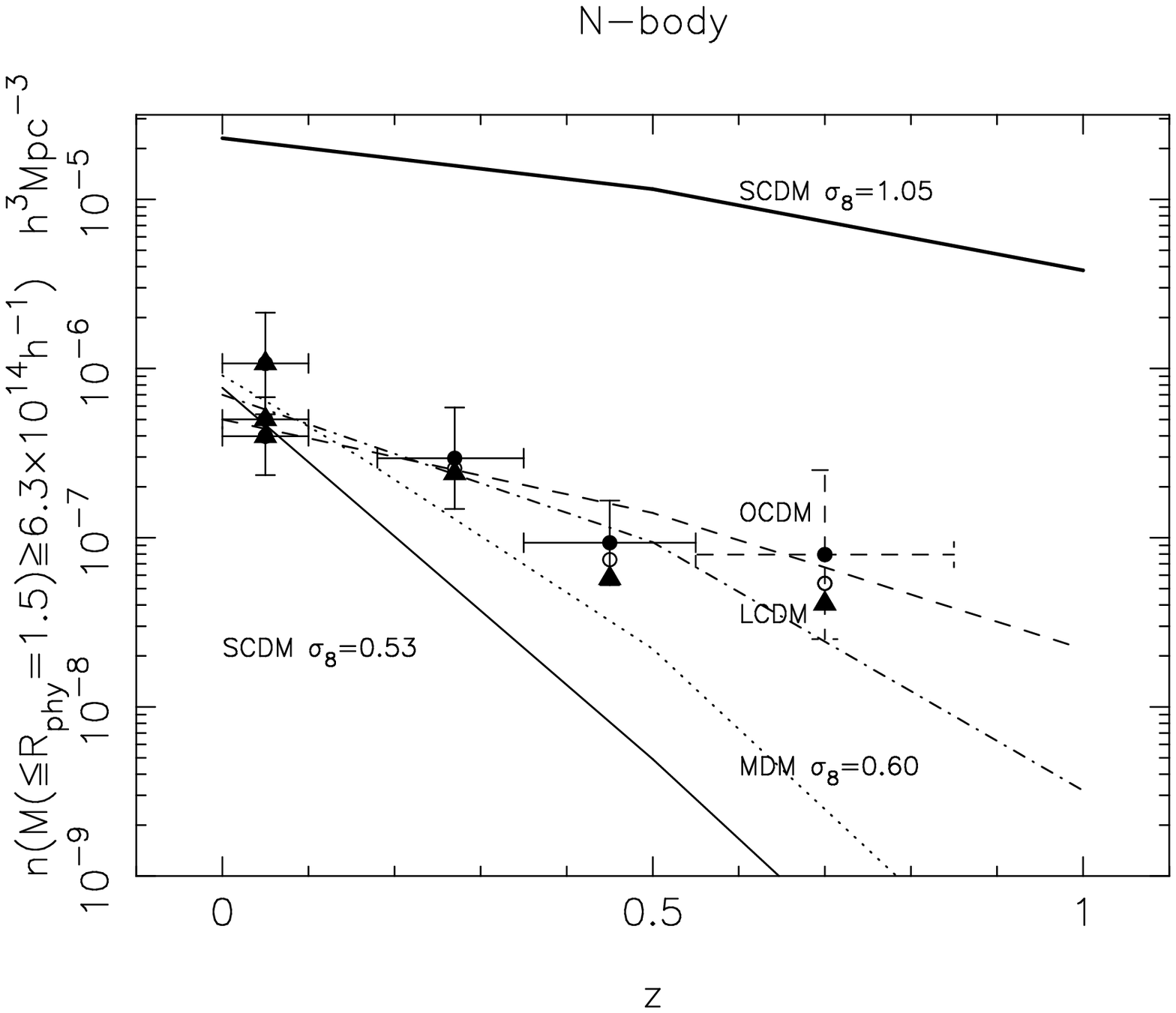}

\vspace{3cm}
Fig. 2 The evolution of cluster abundance with redshift for massive,
coma-like clusters $(M_{1.5}\geq 6.3$ $\times 10^{14}h^{-1}M_{\odot })$. The
lines represent model predictions; the data points are from the CNOC survey.
From Bahcall, Fan and Cen (1997) (updated fig.).

\end{figure}
\begin{figure}
\vspace{-4cm}

\epsfysize=600pt \epsfbox{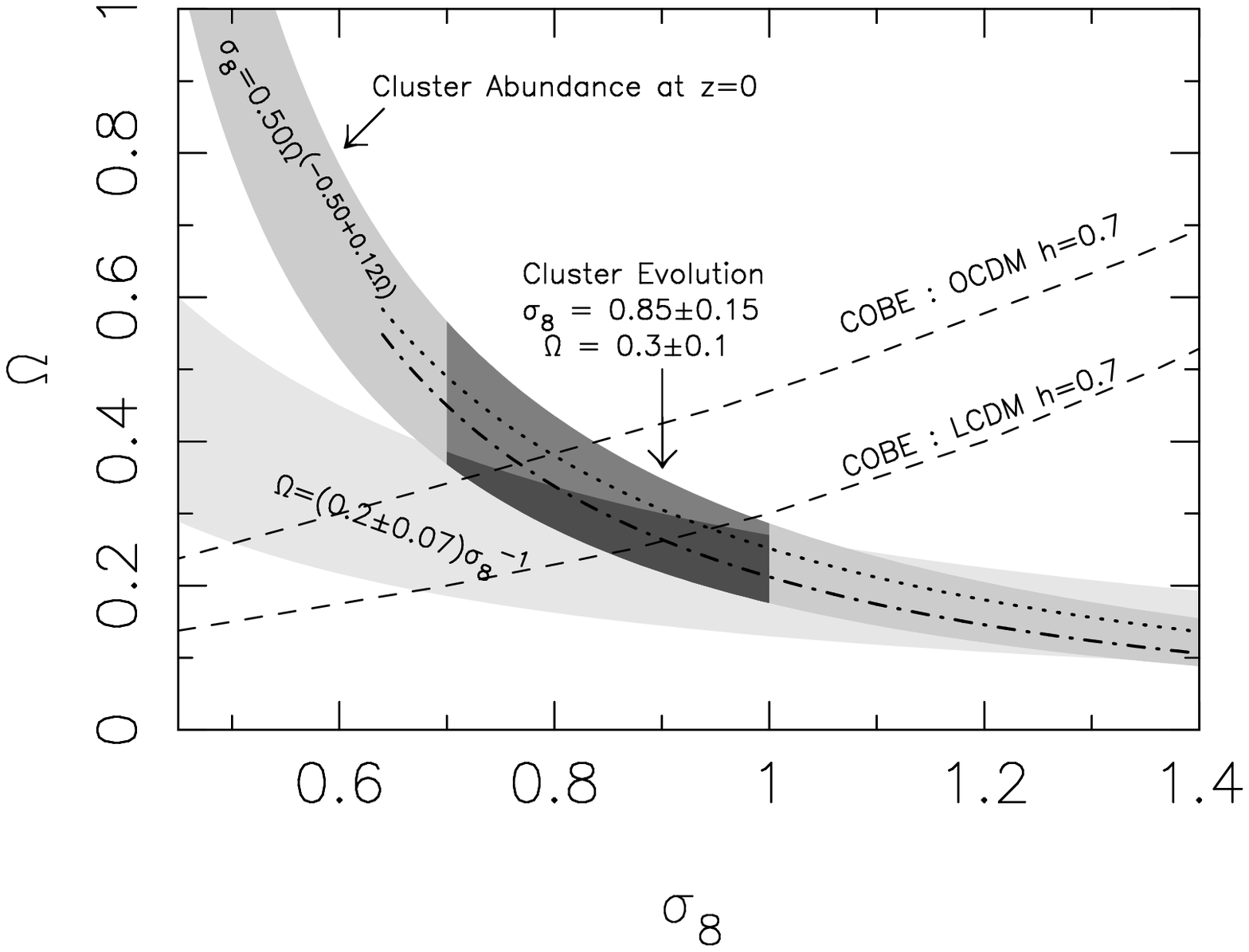}

\vspace{3cm}
Fig. 3 Constraints placed on $\Omega $ and $\sigma _{8}$ by the cluster
evolution data of Fig. 2 (dark region). Other bands indicate the present-day
cluster abundance and the cluster dynamics (Fig. 1) constraints. From
Bahcall, Fan and Cen (1997)
\end{figure}

\newpage
\begin{thebibliography}{99}
\bibitem{}  Bahcall, N.A. 1977, ARA\&A, 15, 505

\bibitem{}  Bahcall, N.A. 1988, ARA\&A, 26, 631

\bibitem{}  Bahcall, N.A. 1995, in AIP conference Proceedings 336, Dark
atter, eds. S.S. Holt \& C.L. Bennett (New York: AIP) p.201

\bibitem{}  Bahcall, N.A. \& Cen, R.Y. 1992, ApJ, 398, L81

\bibitem{}  Bahcall, N.A. \& Cen, R.Y. 1993, ApJ, 407, L49

\bibitem{}  Bahcall, N.A., Fan, X., \& Cen, R. 1997, ApJ, 485, L53

\bibitem{}  Bahcall, N.A., Lubin, L. \& Dorman, V. 1995, ApJ, 447, L81

\bibitem{}  Bahcall, N.A. \& Oh, S.P. 1996, ApJ, 462, L49

\bibitem{}  Bahcall, N.A. \& Soneira, R.M. 1983, ApJ, 270, 20

\bibitem{}  Bahcall, N.A. \& West, M.L. 1992, ApJ, 392, 419

\bibitem{}  Briel, U.G., Henry, J.P. \& Boringer, H. 1992, A\&A, 259, L31

\bibitem{}  Carlberg, R.G., Morris, S.M., Yee, H.K.C., \& Ellingson, E.
1997, ApJ, 479, L19

\bibitem{}  Carlberg, R.G., Yee, H.K.C., Ellingson, E., Abraham, R., Gravel,
P., Morris, S.M., \& Pritchet, C.J. 1996, ApJ, 462, 32

\bibitem{}  Cen, R. \& Ostriker, J.P. 1994, ApJ, 429, 4

\bibitem{}  Croft, R.A.C., Dalton, G.B., Efstathiou, G., Sutherland, W. \&
addox, S.J. 1997, MNRAS, submitted

\bibitem{}  Einasto, J.,{\em et al.}, 1997, Nature, 385, 139; also this
volume

\bibitem{}  Eke, V.R., Cole, S., \& Frenk, C.S. 1996, MNRAS, 282, 263

\bibitem{}  Evrards, A.E., Metzler, C.A., \& Navarro, J.F. 1996, ApJ, 469,
494

\bibitem{}  Fan, X., Bahcall, N.A., \& Cen, R. 1997, ApJ, 490

\bibitem{}  Fischer, P. \& Tyson, J.A. 1997, AJ, 114, 14

\bibitem{}  Jing, Y.P. \& Fang, L.-Z. 1994, ApJ, 432, 438

\bibitem{}  Hughes, J.P. 1989, ApJ, 337, 212

\bibitem{}  Jones, C. \& Forman, W. 1984, ApJ, 276, 385

\bibitem{}  Kaiser, N. \& Squires, G. 1993, ApJ, 404, 441

\bibitem{}  Kitayama, T. \& Suto, Y. 1997, ApJ, in press

\bibitem{}  Klypin, A.A. \& Kopylov, A.I. 1983, Sov. Astron. Lett., 9, 41

\bibitem{}  Lubin, L. \& Bahcall, N.A. 1993, ApJ, 415, L17

\bibitem{}  Lubin, L., Cen, R., Bahcall, N.A. \& Ostriker, J.P. 1996, ApJ,
460, 10

\bibitem{}  Luppino, G.A. \& Gioia, I.M. 1995, ApJ, 445, L77

\bibitem{}  Maddox, S., Efstathiou, G., Sutherland, W. \& Loveday, J. 1990,
NRAS, 242, 43

\bibitem{}  Oukbir, J. \& Blanchard, A. 1997, A\&A, 317, 1

\bibitem{}  Peacock, J. \& Dodds, S.J. 1994, MNRAS, 267, 1020

\bibitem{}  Peebles, P.J.E. 1980, The Large-Scale Structure of the Universe
(Princeton: Princeton Univ. Press)

\bibitem{}  Peebles, P.J.E. 1993, Principles of Physical Cosmology
(Princeton: Princeton Univ. Press)

\bibitem{}  Pen, U.-L. 1997, ApJ, submitted

\bibitem{}  Press, W.H. \& Schechter, P. 1974, ApJ, 187, 425

\bibitem{}  Retzlaff, J., Borgani, S., Gottlober, S. \& Mueller, V. 1997,
NRAS, submitted

\bibitem{}  Sarazin, C.L. 1986,Rev. Mod. Phys., 58, 1

\bibitem{}  Smail, I., Ellis, R.S., Fitchett, M.J., \& Edge, A.C. 1995,
NRAS, 273, 277

\bibitem{}  Strauss, M. \& Willick J 1995, Physics Reports, 261, 271

\bibitem{}  Tadros, H., Efstathiou, G. \& Dalton, G. 1997, MNRAS, submitted

\bibitem{}  Tyson, J.A., Wenk, R.A. \& Valdes, F. 1990, ApJ, 391, L5

\bibitem{}  Viana, P.P. \& Liddle, A.R. 1996, MNRAS, 281, 323

\bibitem{}  White, S.D.M., Navarro, J.F., Evrard, A. \& Frenk, C.S. 1993,
Nature, 366, 429

\bibitem{}  White, S.D.M., Efstathiou, G., \& Frenk, C.S. 1993, MNRAS, 262,
1023

\bibitem{}  White, D. \& Fabian, A. 1995, MNRAS, 273, 72
  
\bibitem{}  Zwicky, F. 1957, Morphological Astronomy (Berlin:
Springer-Verlag).
\end{thebibliography}
\end{document}